\documentclass[12pt]{article}
\usepackage{amsmath,amssymb}

\newcommand{\p}{\bot}
\newcommand{\dd}{\partial}

\newcommand{\e}{\varepsilon}

\newcommand{\m}{\mu}
\newcommand{\n}{\nu}

\newcommand{\disn}[2]{$$\displaylines{\refstepcounter{equation}%
            \label{#1}\hskip 1em minus 1em #2\hfilneg}$$}
\newcommand{\nom}{\hfil\hskip 1em minus 1em (\theequation)}
\newcommand{\no}{\hfil \hskip 1em minus 1em\phantom{(\theequation)}%
            \hfilneg\cr\hfilneg\hskip 1em minus 1em\hfil}

\begin{document}

\title{\bf
Gauge invariant regularization of QCD on the Light Front with the
lattice in transverse space coordinates}

\author{\bf M.Yu. Malyshev\thanks{E-mail: mimalysh@yandex.ru}, E.V. Prokhvatilov
\\\text{Saint Petersburg State University, Saint Petersburg, Russia}}
\date{}
\maketitle

\begin{abstract}
It is introduced the gauge invariant regularization of quantum
chromodynamics (QCD), adjusted to modeling nonperturbative
vacuum effects
in QCD  on the  light front (LF) via modeling the dynamics
of zero Fourier modes of fields on the LF.
\end{abstract}

\section{Introduction}
Quantization of field theory on the light front (LF) \cite{dir},\,\,
 i.e. on the hyperplane  $x^+=0$ in
the coordinates  $$x^{\pm}=(x^0\pm x^3)/\sqrt{2},\quad x^{\p}=x^1,x^2,$$
where $x^0,x^1,x^2,x^3$ are Lorentz coordinates and the $x^+$ plays
the role of
the time, requires special regularization of the theory.
The LF momentum operator $P_-$
(the generator of translations along
the $x^-$ axis) is nonnegative for the states with nonnegative
energy and mass:
$$P_-=(P_0-P_3)/\sqrt{2}\geqslant 0\quad  \text{for} \quad p_0\geqslant0,\, p^2\geqslant0.$$
The vicinity of its minimal eigenvalue,
$p_-=0$, corresponds both to ultraviolet and infrared domains of
momenta in Lorentz coordinates. Quantizing field theory on the LF one
finds singularities at $p_-\to 0$.
And the regularization of these singularities may affect
the description of both
ultraviolet and infrared momenta physics, in particular,
correct description of vacuum effects.

Usual ways of the regularization of the $p_-\to 0$ singularities
are the following:

{\bf (a)} the cutoff $|p_-|\;\; (|p_-|\geqslant\e>0)$,

{\bf (b)}  "DLCQ"  regularization, i.e. the space cutoff in the
 $x^-$, $|x^-|\leqslant L$, plus the periodic boundary conditions on fields $x^-$,
 that leads to the discretization of the $P_-$ spectrum:
 $p_-=\frac{\pi n}{L}$, $n=0,1,2,\dots$
The Fourier mode of the field with the $p_-=0$ ("zero mode")
is separared here from other modes.
In canonical formalism zero mode turns out to be dependent on other modes
 due to constraints
(for gauge field theory see \cite{nov2,nov2a}).

Both ways of the regularization break Lorentz symmetry, and the
regularization (a)
violates also gauge invariance in gauge field theory.
This can lead to difficulties with the renormalization of
the theory, and also to
a nonequivalence of results, obtained with LF and with usual
("equal time") quantization.
In the framework of perturbation theory it was shown
\cite{burlang, tmf97}  that to restore the symmetry and the
above mentioned equivalence it is necessary to add to the regularized
LF Hamiltonian
some special "counterterms".

However for the Quantum Chromodynamics (QCD) one can
expect effects nonperturbative in coupling
 constant,
in particular, vacuum condensates. Applying the regularization of the
type (a) where one excludes zero
modes, we get
the absence of such condensates. The regularization (b) leads
to canonically
constrained and dynamically not independent zero modes. With
these zero modes
 one again cannot correctly
descibe condensates \cite{yaf88, yaf89}. The study of this problem in
(1+1)-dimensional quantum electrodynamics suggested some way to introduce
 correct description of condensates in the regularization (b), at
least semiphenomenologically,
using zero modes as independent variables \cite{yaf88, yaf89}.

In the present paper we review briefly our new parametrization of gauge fields
on the lattice in "transversal" space coordinates  on the LF. This parametrization is
convinient for separate treatment of zero modes of fields on the LF and gives a  way to introduce
gauge invariant regularization of the theory. Then we limit ourselves  by QCD(2+1) model in coordinates
close to the LF and perform the limiting transition to the LF Hamiltonian keeping
the dynamical independence of zero modes of fields. We apply this Hamiltonian for simple
example of mass spectrum calculation.

\section{The definition of gauge fields on the "transverse" lattice}

The gluon part of QCD Lagrangian in continuous space has the following form:
\disn{1}{
{\cal L}=-\frac{1}{2} Tr F_{\m\n}F^{\m\n}.
\nom}
where
$$F_{\mu\nu}= \dd_{\mu} A_{\nu}- \dd_{\nu }A_{\mu}-ig[A_{\m},A_{\n}]$$
and the gluon vector fields $A_{\mu}(x)$
are $N\times N$ Hermitian traceless matrices. Under SU(N) gauge transformations
the $A_{\mu}(x)$ transform as follows:
\disn{2}{
A_{\mu}(x) \to \Omega(x)A_{\mu}(x)\Omega^+(x)+ \frac {i}{g}
\Omega(x)\dd_{\mu}\Omega^+(x).
\nom}
Here the $\Omega (x)$ are  $N\times N$ matrices, corresponding
to the SU(N) gauge transformation.

In the LF Hamiltonian approach one uses continuous  coordinates $x^+, x^-$
and inroduces, as an ultraviolet
regulator, the lattice in transversal coordinates.
Gauge invariance is maintained via appropriate use of Wilson lattice method
\cite{wilson}, describing gauge fields by matrices related to lattice links.

If one uses the unitary matrices for these link variables  and  constructs
the Hamiltonian, one needs to apply the "transfer matrix" method, described in the paper \cite{creutz, Grunewald}.
However this method is not accomodated to the LF and to the corresponding choice
 of the gauge $A_-=0$.

To overcome this difficulty we propose the modification of these link variables,
introducing
nonunitary matrics of special form, where only zero modes are related with
links and nonzero modes are related with the sites, belonging to these links.
Using these lattice variables we can represent the complete regularization of the
theory in gauge invariant form.

The gluon field components
$A_+$ and $A_-$ are related with the lattice sites. Under the gauge transformations
they transform according to previous formulae (\ref{2}).
Transverse components are described by the following $N\times N$ complex matrices:
\disn{3}{
M_{\m}(x)=(I+iga\tilde A_{\m}(x))U_{\m}(x),
\nom}
where $m$ is the index of transversal components and the $\tilde A_{\m}(x)$ are
Hermitian $N\times N$ matrices,\,
related to corresponding lattice sites, $U_{\m}(x)$ are unitary
$N\times N$ matrices, related to the links $(x-ae_{\m}, x)$,
$a$ is the parameter of the lattice (the size of the link) and the $e_{\m}$ is
the unit vector along the $x^{\m}$ axis,
$g$ is the QCD coupling constant.

We define the transformation law under gauge transformations as follows:
\disn{4}{
\tilde A_{\mu}(x) \to \Omega(x)\tilde A_{\mu}(x)\Omega^+(x),\quad
U_{\m}(x)\to \Omega(x)U_{\m}(x)\Omega^+(x-ae_{\m}) .
\nom}

In consequence the matrices $M_{\m}(x)$ transform like link variables \cite{lat,lat1}:
$$M_{\m}(x)\to \Omega(x)M_{\m}(x)\Omega^+(x-ae_{\m}).$$

Let us remark that the Hermicity of matrices $\tilde A_{\m}(x)$ is kept under
these gauge transformations.

Let us introduce the operator $D_-$ by the following definitions:
\disn{5}{
D_-\tilde A_{\m}(x)=\dd_-\tilde A_{\m}(x)-ig[A_-(x),\tilde A_{\m}(x) ],\no
D_-U_{\m}(x)=\dd_-U_{\m}(x)-igA_-(x) U_{\m}(x)+igU_{\m}(x) A_-(x-ae_{\m}),\no
D_-M_{\m}(x)=\dd_-M_{\m}(x)-igA_-(x) M_{\m}(x)+igM_{\m}(x) A_-(x-ae_{\m}).
\nom}
This definition of the $D_-$ has gauge invariant form under the gauge transformations,
defined above.

Further we impose on the $U_{\m}(x)$ the condition
\disn{6}{
D_-U_{\m}(x)=0,
\nom}
while from the $\tilde A_{\m}(x)$ we exclude the part, which satisfies the equality
$D_-\,\tilde A_{\m}(x)=0$. In the gauge $A_-=0$ these conditions simply
mean a
separation of zero ($U_{\m}(x)$) and nonzero ($\tilde A_{\m}(x)$)
Fourier modes of
the field in the $x^-$. In general we have some
gauge invariant definition of this
separation.

Furthermore we can introduce
the gauge invariant cutoff in $p_-$, using a cutoff \,
in the eigenvalues $q_-$ of the $D_-$: $|q_-|\leqslant \Lambda$.

Now let us consider the naive continuous space limit $a\to 0$. We require
the following relation in the fixed gauge $A_-=0$ at $a\to 0$:
$$U_{\m}(x)\to \exp{igaA_{\m 0}(x)}\to (I + igaA_{\m 0}(x)),$$
Here the $A_{\m 0}(x)$ is zero mode of
the field $A_{\m}(x)$ in continuous
 space. And for the $\tilde A_{\m}(x)$ we require that it tend
  to nonzero mode part of the $A_{\m}(x)$.
Then at nonzero $A_-$ we can get for the matrix $M_{\m}(x)$
the following relation:
\disn{7}{
M_{\m}(x)\to (I+iga A_{\m}(x)+ O((ag)^2)).
\nom}

Indeed, at $a\to 0$ we have:
\disn{8}{
M_{\m}(x)\to\Omega(x; A_-)(I+iga A_{\m}(x))_{A_-=0}
\Omega^+(x-ae_{\m}; A_-)\to\no
\to \Omega(x; A_-)(I+iga A_{\m}(x))_{A_-=0}\Omega^+(x; A_-)-a
\Omega(x; A_-)\dd_{\m}\Omega^+(x; A_-)\no
\to (I+iga A_{\m}(x)),
\nom}
where the $\Omega (x; A_-)$ is the matrix of the gauge transformation,
which transforms the field in the gauge $A_-(x)=0$ to the field with a
given $A_-(x)$.

Let us introduce the lattice analog of the continuous space field strength
 $F_{\m\n}(x)$. With this aim we define the following quantities
 ($\m, \n = 1,2$):
\disn{9}{
  G_{\m\n}(x)=-\frac
{1}{ga^2}[M_{\m}(x)M_{\n}(x-ae_{\m})-M_{\n}(x)M_{\m}(x-ae_{\n})],
\nom}
\disn{10}{
 G_{+-}(x)=iF_{+-}(x),\quad G_{-\m}= \frac{1}{ga}D_-M_{\m},\no
 G_{+\m}(x)=\frac
  {1}{ga}[\dd_{+}M_{\m}(x)-ig(A_{+}(x)M_{\m}(x)
  -M_{\m}(x)A_{+}(x-ae_{\m}))].
\nom}

 It is not difficult to show that at $a\to 0$ one gets
   $G_{\m\n}(x)\to iF_{\m\n}(x)$, and the analogous relations  are
   true for the
 $G_{+\m}$, $G_{-\m}$.

We get the following transformation law under the gauge transformations:
\disn{11}{
G_{\pm\m}(x) \to \Omega(x)\,G_{\pm\m}(x)\,\Omega^+(x-ae_{\m}),\no
G_{\m\n}(x)\to \Omega(x)\,G_{\m\n}(x)\,\Omega^+(x-ae_{\m}-ae_{\n}).
\nom}

Having these quantities one can construct
gauge-invariantly regularized action and
the LF Hamiltonian of the QCD (similarly to the work \cite{lat1}).

One can also apply the transfer matrix method of the paper \cite{creutz}
to construct the Hamiltonian on the
lattice even in the gauge $A_-=0$, because  only zero modes are
 described
by unitary matrices on links, and it is possible
to find the necessary
"transfer matrix" in $x^+$, in analogy with paper \cite{creutz}.
\vskip 1em
{\bf Acknowledgements.}
We thank V.A.~Franke and S.A.~Paston for useful discussions.

\end{document}